\documentclass{elsart}

\usepackage{graphicx}

\usepackage{amsfonts,amssymb,amsmath}
\usepackage{multirow}

\usepackage{natbib}
\begin{document}
\maketitle

\begin{frontmatter}



\title{A Dynamical Model of Lipoprotein Metabolism}


\author{E.~August, K.H.~Parker and M.~Barahona\corauthref{cor1}}
\corauth[cor1]{Corresponding author}

\address{Department of Bioengineering, Imperial College London,\\
South Kensington Campus, London SW7 2AZ, United Kingdom}

\begin{abstract}
We present a dynamical model of lipoprotein metabolism derived by
combining a cascading process in the blood stream and cellular
level regulatory dynamics. We analyse the existence and stability
of equilibria and show that this low-dimensional, nonlinear model
exhibits bistability between a low and a high cholesterol state. A
sensitivity analysis indicates that the intracellular
concentration of cholesterol is robust to parametric variations
while the plasma cholesterol can vary widely. We show how the
dynamical response to time-dependent inputs can be used to
diagnose the state of the system. We also establish the connection
between parameters in the system and medical and genetic
conditions.
\end{abstract}

\begin{keyword}
lipoprotein metabolism  \sep dynamical systems \sep nonlinear
models \sep metabolic control mechanisms  \sep statins \sep
familial hypercholesterolemia
\end{keyword}

\end{frontmatter}


\section{Introduction}
\label{Intro}

We model the dynamics of lipoprotein metabolism associated with
the transport of lipids in the blood. The malfunction of the
delivery of lipids and cholesterol from the liver to the cells has
important ramifications. There is strong medical evidence linking
high plasma concentrations of low-density lipoproteins (LDL) to
the development of atherosclerosis, the deadliest disease in
industrialised countries~\citep{GW01,Rodriguez99,PL02}. However,
LDL is just one species in a metabolic cascade by which different
lipoproteins~(Table~\ref{Table0}) are synthesised, degraded and
absorbed. In this complex network, it is important to investigate
how metabolic control is exercised and if dynamical features,
other than high LDL concentrations, could be useful markers for
the detection of disease states.

Previous modelling efforts of lipoprotein metabolism vary in
scope. Some models have concentrated on detailed aspects of the
system, such as the fluid dynamics of lipid accumulation on the
arterial walls~\citep{ALHTJP98} or the chemical kinetics of LDL
oxidation~\citep{CAC02}, but do not describe the processes at the
cellular or physiological level. Other approaches have modelled
the lipoprotein network as a whole with linear compartmental
models~\citep{JAJ85,FP98,
KGP91}. Although these models can be fitted to match experimental
data, the compartments lack rigorous physiological meaning, thus
making it difficult to describe the underlying biochemical
processes. For instance, compartments added to represent
subclasses of lipoproteins can always be reduced to one
compartment because of the linearity of the
models~\citep{JESMG84}.

Our model of lipoprotein metabolism is a relatively
low-dimensional system of nonlinear differential equations which
are strongly linked to the underlying physiological processes. The
system can undergo a transition between low and high LDL steady
states, a feature which is robust to randomness and uncertainty in
the parameters. Because the equations are directly related to
physiological processes, the model can be used to study the effect
of genetic or behavioral conditions and could serve as an aid to
test hypotheses for diagnosis and intervention. In what follows,
we derive our model from the main physiological processes; we then
analyse the system and carry out a sensitivity analysis; finally,
we show how the results can be related to experiments and medical
conditions.

\section{Modelling Lipoprotein Metabolism}\label{modelling}

\begin{figure}
\vspace*{-1.75cm}
\centering
\includegraphics[width=0.85 \textwidth]{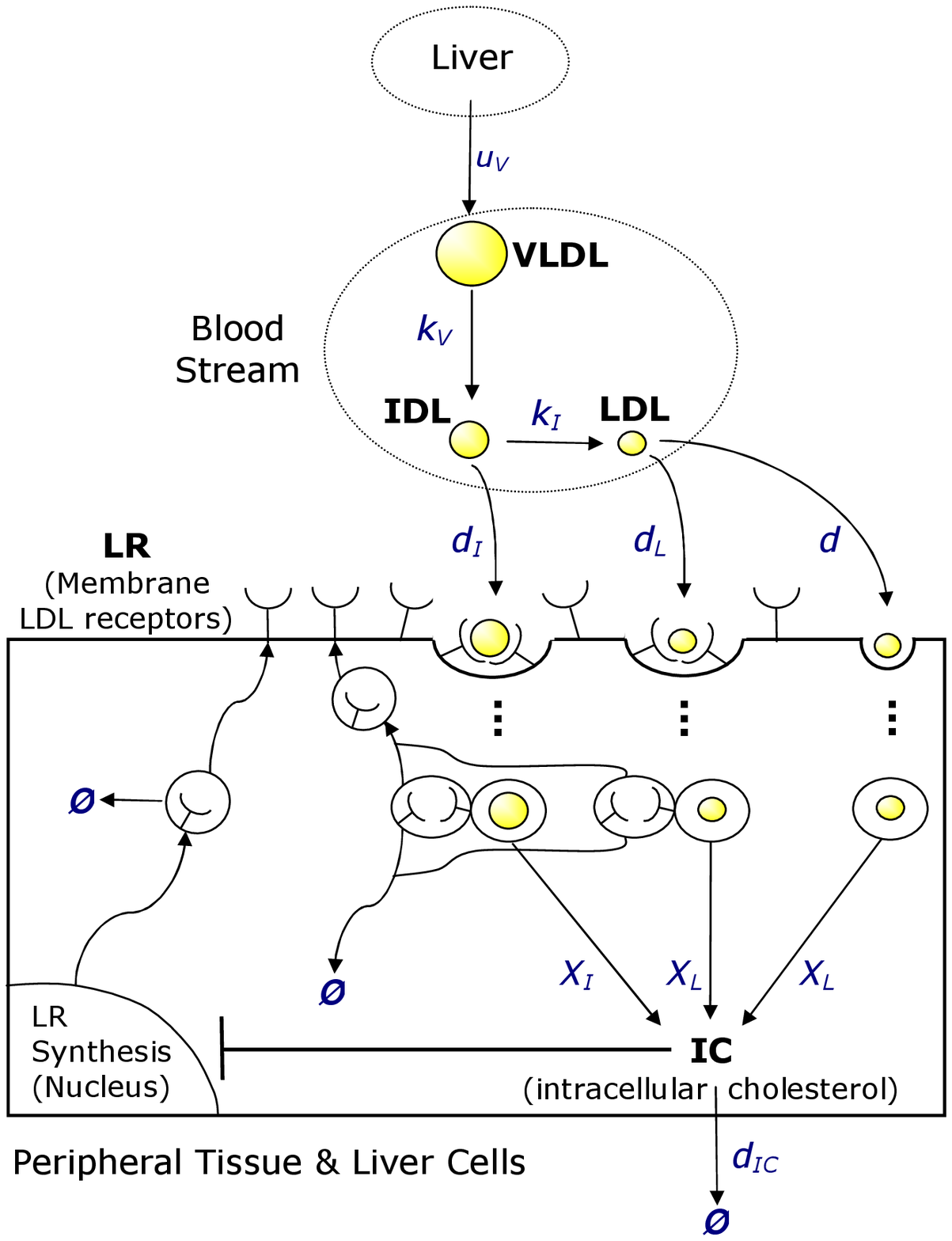}
\caption[\textbf{Overview of origin, transport and fate of
lipoproteins.}]{\textbf{Overview of origin, transport and fate of
lipoproteins.} {\small A schematic view of the processes involved
in the metabolism of lipoproteins at both the macroscopic and
cellular levels. The liver secretes VLDL at a rate $u_V$. The
turnover rates of VLDL and IDL in the blood stream are $k_V$ and
$k_I$, respectively. At the cellular level,  $d_L$ and $d_I$
denote rate constants of the receptor-mediated uptake of
lipoproteins: IDL and LDL bind to membrane receptors (LR), which
cluster in coated pits in the cell membrane. Note that it is also
possible for LDL (but not for IDL) to be absorbed directly through
a non-receptor mediated path with rate constant $d$. After
endocytic vesicles are formed and internalised, the lipoproteins
are hydrolysed in the cell releasing lipids into the cytoplasm. A
significant proportion of these lipids ($\chi_I$ in IDL and
$\chi_L$ in LDL) is cholesterol which contributes to the level of
intracellular cholesterol (IC). The intracellular cholesterol is
used for cell function or eliminated, mainly through the action of
HDL, at a rate $d_{IC}$. Most of the LDL receptors (but not all)
are reincorporated into the membrane. In addition, the nuclear
synthesis of the receptors LR is regulated through negative
feedback ($\vdash$) by the intracellular cholesterol
IC.}}\label{fig7}
\end{figure}

Lipoproteins constitute the primary means of transport of lipids
from the liver to the cells via the blood stream. Lipoprotein
dynamics is intimately connected to lipid and cholesterol
metabolic networks. 
Figure~\ref{fig7} provides a schematic overview of the origin,
transport and fate of the different lipoproteins in the human body
at two levels: the macroscopic and the cellular.

Lipoproteins are globular aggregates of lipids and proteins:
cholesterol ester and triacylglycerol form the core; amphiphilic
phospholipids surround them forming the hull; and different
apolipoproteins (apo A, apo B, apo C, and apo E), which function
as trigger-molecules for specific reactions, are embedded in the
surface. Lipoproteins are classified in five standard groups
(Table~\ref{Table0}), based on density and the apolipoproteins
attached to them. In increasing order of density and decreasing
order of size (the fewer lipids relative to its size, the denser
the lipoprotein is): chylomicrons (Chyl), very low-density
lipoproteins (VLDL), intermediate-density lipoproteins (IDL),
low-density lipoproteins (LDL), and high-density lipoprotein
(HDL).

\begin{table}[htbp]
\begin{tabular}{|l|l|l|l|l|l|}
\hline & Chyl & VLDL & IDL & LDL & HDL\\
\hline
Density ($g/ml$)  & $<$ 0.95 &  0.95--1.006 &  1.006--1.019 &  1.019--1.063 &  1.063--1.21\\
Diameter ($nm$) &  80--100 & 30--80 &  25--30 &   20--25 &  8--13\\
Cholesterol (\%) &  3--5 &   12--21 &   27--46 &   40--50 &   15--25\\
Apolipoproteins& A, C, E, B-48 &  C, E, B-100 &  C, E, B-100 &  B-100 &  A, C, E\\
\hline
\end{tabular}\\ \caption{\textbf{The properties and
composition of the five standard lipoprotein classes}
\citep{MA02}: chylomicrons (Chyl), very low-density lipoproteins
(VLDL), intermediate-density lipoproteins (IDL), low-density
lipoproteins (LDL), and high-density lipoproteins
(HDL).}\label{Table0}
\end{table}

Chylomicrons contain a large proportion of fatty acids and are
found in the blood stream mainly after digestion of a meal. They
are synthesised by the intestines and constitute the direct means
by which dietary fat is delivered to heart, muscle, and adipose
tissue. Chylomicrons provide the liver with lipids and proteins
but are not regulated by the liver. Therefore, we do not consider
the concentration of chylomicrons as an explicit variable in our
model and include their effect parametrically as an input of
lipids to the system.

The liver is responsible for the secretion of all other
lipoproteins, most significantly VLDL. (Although small amounts of
IDL and LDL are produced by the liver, we neglect this
contribution.) VLDL is secreted by the liver into the blood at an
almost constant rate (slightly higher during the night) as a means
of transport of synthesised fat and cholesterol to peripheral
tissues. While in the blood stream, the apo C protein on the
surface of VLDL can bind to the extracellular enzyme lipoprotein
lipase (LPL), which is present on the capillary wall. The effect
of LPL is to hydrolyse triacylglycerol, thus reducing the size of
VLDL and increasing the percentage of denser molecules (i.e., it
degrades VLDL to IDL). After hydrolysis, the fatty acids are taken
up by nearby cells or by serum albumin for transport to more
peripheral cells. A similar process leads from IDL to LDL. This is
the lipoprotein cascade that transforms VLDL into IDL into LDL.
(This process also involves losing apo C and apo A to HDL, which
is independently synthesised in the liver and is not a part of the
lipoprotein cascade.)

About 50\% of the IDL and 75\% of LDL in the blood are absorbed by
hepatic and peripheral cells via LDL receptors (LR) on the cell
membrane. These receptors are synthesised by the cell and
recognise apo B and apo E with high affinity. In liver cells, the
absorbed LDL is reused for lipoprotein synthesis and excess
cholesterol is secreted into bile. In non-hepatic cells, the
absorbed LDL supplies the cholesterol content essential for cell
function~\citep{DAW78, MCDIT99}. Although almost every cell can
synthesise cholesterol to some extent, we assume that all
cholesterol has to be delivered to the cells via LDL or IDL
absorption~\citep{CKM00,GMC00,CACERS92,JMD78}.

As mentioned above, there is another important component in the
lipoprotein network: HDL (the ``good cholesterol''), which is also
secreted by the liver. HDL is responsible for \emph{reverse
cholesterol transport}, the transport of excess cholesterol from
cells and other lipoproteins back to the liver. The effect of HDL
is included in our model through a rate constant that
parameterises the elimination of intracellular cholesterol, which
is significantly affected by the HDL pathway.

The above processes can be modelled with the following system of
differential equations, where $[\cdot]$ denotes concentration and
$\phi_{LR}$ is the dimensionless fraction of total possible LDL
receptors:
\begin{align}
\frac{d[VLDL]}{dt}&=-k_V[VLDL]+u_{V}\label{full1}\\
\frac{d[IDL]}{dt}&=k_V[VLDL]-k_I[IDL]-d_I[IDL]\phi_{LR}\label{full2}\\
\frac{d[LDL]}{dt}&=k_I[IDL]-d_L[LDL]\phi_{LR}-d[LDL]\label{full3}\\
\frac{d\phi_{LR}}{dt}&=-b \left(d_I[IDL]+d_L[LDL] \right )\phi_{LR}+c\frac{1-\phi_{LR}}{[IC]}\label{full4}\\
\frac{d[IC]}{dt}&=\left(\chi_Id_I[IDL]+\chi_Ld_L[LDL]\right)\phi_{LR}+\chi_Ld[LDL]-d_{IC}[IC]
.\label{full5}
\end{align}
The linear terms $k_V [VLDL]$ and $k_I [IDL]$
in~(\ref{full1})--(\ref{full3}) correspond to the cascade
degradation of VLDL to IDL to LDL assuming saturated kinetics
(i.e., excess of LPL enzyme) for the reactions. An implicit
assumption is that the blood flow does not play a significant role
in the absorption process. This is based on our analysis of the
relevant time scales and also on detailed calculations of mass
transfer in tissue under physiological
conditions~\citep{RossEthMasters05}.

Equation~(\ref{full1}) contains the only input to the system,
$u_V$, which represents the overall secretion rate of VLDL by the
liver. High values of $u_V$ are associated with a high dietary
intake of fats~\citep{DKS93}. On the other hand, medication with
statins would reduce $u_V$ by lowering the synthesis of
cholesterol and VLDL in the liver.

The nonlinear terms $d_I[IDL]\phi_{LR}$ and $d_L[LDL]\phi_{LR}$
represent the endocytosis of IDL and LDL  via a process involving
LDL receptors (LR) and leading to the release of lipids and
cholesterol into the cytoplasm. This can be modelled in its
simplest, classical form through a second order
reaction~~\citep{RPB02}. In addition to this receptor-mediated
uptake, mammalian cells can take up LDL (but not IDL) by
non-specific endocytosis at a rate $d[LDL]$, linearly proportional
to its plasma (or extracellular) concentration
\citep{HHH90,TEW99,GW01,GB97}. Endocytosis is greatly reduced
under the genetic disease {\it familial
hypercholesterolemia}~(Section~\ref{FH}).

The last two equations~(\ref{full4})--(\ref{full5}) describe
cholesterol uptake and regulation at the cellular level. The
variables are: $\phi_{LR}$, the fraction of the total possible LDL
receptors that mediate IDL and LDL endocytoses; and $[IC]$, the
concentration of cellular cholesterol, which closes the loop by
acting as a cellular control for the expression and synthesis of
the LDL receptors, LR. {\small
\begin{table}[h]
\begin{tabular}{|c|c|c|c|c|}
\hline
Parameter & Source & Units & Nominal Value & Range\\
\hline
$k_V$ & \citep{CJP00} & $h^{-1}$ &$0.3$ & 0.15--0.6\\
$k_I$ & \citep{CJP00} & $h^{-1}$ &$0.1$ & 0.025--0.1\\
$d_I$ & --- & $h^{-1}$ & $1.4$ & 0.5--2\\
$d_L$ & --- & $h^{-1}$ &$0.005$ & 0.005--0.02\\
$d$ & \citep{JMD93} & $h^{-1}$ & 0.0025& 0.0025-0.0075\\
$b$ & --- & $lg^{-1}$ & $0.1$& 0--2 \\
$c$ & {\small \citep{JM77}} & $g(lh)^{-1}$ & $0.05$ & 0--1\\
$\chi_I$ & \citep{MA02} & -- & 0.35 & 0.25--0.45\\
 $\chi_L$ &  \citep{MA02} & -- & 0.45 & 0.4--0.5\\
\hline
$u_{V}$ & \citep{DAW78} & $g(lh)^{-1}$ & $0.3$ & Variable\\
$d_{IC}$ & \citep{DAW78} & $h^{-1}$ & $0.45$ & Variable\\ \hline
\end{tabular} \\
\caption{\textbf{Parameters of the
model~(\ref{full1})-(\ref{full5}).} {\small Nominal values and
ranges as found in the literature. The nominal values for $d_I$
and $d_L$ are obtained by assuming that all other parameters are
at nominal values and the equilibrium point of the linearised
system in the lower branch of Eq.~(\ref{eq:linearasymptotic}) is
that of normal individuals~\citep{DAW78}: $[VLDL]=1 \, gl^{-1}$,
$[IDL]=0.2 \,gl^{-1}$, $[LDL]=2 \,gl^{-1}$. The nominal values for
$u_V$ and $d_{IC}$ are crude estimates from the cited reference.
Because of their high dependence on diet, medication and genetic
factors, we consider $u_V$ and $d_{IC}$ as the control
parameters.}} \label{Table1}
\end{table}}

Eq.~(\ref{full4}) reflects the dynamical processes involving the
LDL \emph{surface} receptors, which take part in the endocytotic
reactions. Experiments in normal individuals indicate that the
maximum number of receptors on the surface of the cell is $15,000
- 70,000$ per cell~\citep{JM77}. The variable $\phi_{LR}$
represents the fraction of the total possible surface receptors
that are present on the surface of the cell. Therefore,
$\phi_{LR}$ is a surface occupancy fraction; hence it is
dimensionless and bounded ($0 \leq \phi_{LR} \leq 1$). Biological
evidence indicates that most (but not all) of the receptors used
in the endocytosis are reintegrated into the cell membrane. The
first term in~(\ref{full4}), weighted by a factor $b$, represents
the loss of LDL receptors that are \emph{not} recycled. The last
term in~(\ref{full4}) describes the cholesterol-regulated
synthesis of LDL receptors and the process of their attachment to
the surface of the cell. Further biological evidence indicates
that cells regulate the synthesis of LDL receptors to prevent the
over-accumulation of intracellular cholesterol. When grown in the
presence of different concentrations of LDL, cells adjust the
number of LDL receptors to take up only enough LDL to satisfy the
cholesterol requirement for membrane turnover~\citep{JM77}; when
grown in total absence of LDL, the maximum number of possible
surface LDL receptors is reached after 2--3 days. We have modelled
this inhibitory mechanism with a simple reciprocal feedback: the
synthesis of cytosolic LDL receptors is inversely proportional to
the intracellular cholesterol $[IC]$. Furthermore, the attachment
of the cytosolic LDL receptors to the cell surface can be
described in its simplest form with a rate that is proportional to
the fraction of unoccupied receptor sites $(1-\phi_{LR})$. The
combined rate at which $\phi_{LR}$ increases is then given by the
product $c (1-\phi_{LR})/[IC]$, where $c$ modulates the weight of
the combined process of regulated synthesis and attachment.
However, we emphasise that the qualitative results of the model do
not depend on the particular form of the feedback chosen, as we
discuss at the end of Section~\ref{sens}. For a more detailed
explanation of this term see Appendix~\ref{appendix1}.

Finally, Eq.~(\ref{full5}) establishes the balance of flows of the
cellular cholesterol $[IC]$. The parameters $0< \chi_I < \chi_L <
1$ are the proportions of cholesterol in IDL and LDL,
respectively, which can be measured experimentally. The last term
in the equation $d_{IC}[IC]$ represents the outflow rate of
cholesterol from the cell. This is largely related to the
HDL-mediated \emph{reverse cholesterol transport}, but also
accounts for the cholesterol used by cells for their metabolism or
eliminated through bile secretion in the liver. Therefore, our
model parameterises the effect of HDL through the constant
$d_{IC}$. As in the case of $u_V$, $d_{IC}$ will be a highly
variable control parameter which can be affected by diet and
statins.

The model has eleven parameters with nominal values and ranges
shown in Table~\ref{Table1}:
\begin{description}
\item {\textbf{Model parameters:}} $k_V,\ k_I,\ d_I,\ d_L,\ d,\
b,\ c,\ \chi_I$, and $\chi_L$. These will differ between
individuals but are assumed to remain virtually unchanged for an
individual except for slow changes (e.g., with age). We have found
values and estimates in the literature for all of these
parameters, except for $b$ for which we assume $b=0.1$.
Sensitivity analyses will be performed 
to check that the qualitative results are robust to variations in
these parameters.

\item {\textbf{Control parameters:}} $u_V$ and $d_{IC}$. The
secretion rate of VLDL from the liver ($u_V$) and the depletion
rate of intracellular cholesterol ($d_{IC}$) are highly variable
parameters and affected by genetic factors, diet and medication.
Hence, we will consider these as the control parameters for our
analysis.
\end{description}

\section{Model Analysis: Fixed Points and Sensitivity Analysis}\label{section:modelanalysis}

\subsection{The System at Equilibrium}\label{modelequilibrium}

We have carried out a numerical analysis of the bifurcations and
fixed points of the dynamical system (\ref{full1})--(\ref{full5}).
Before presenting the numerics, we briefly note some analytical
features of the model.

\begin{figure}[h]
\centering
\includegraphics[width=0.95\textwidth]{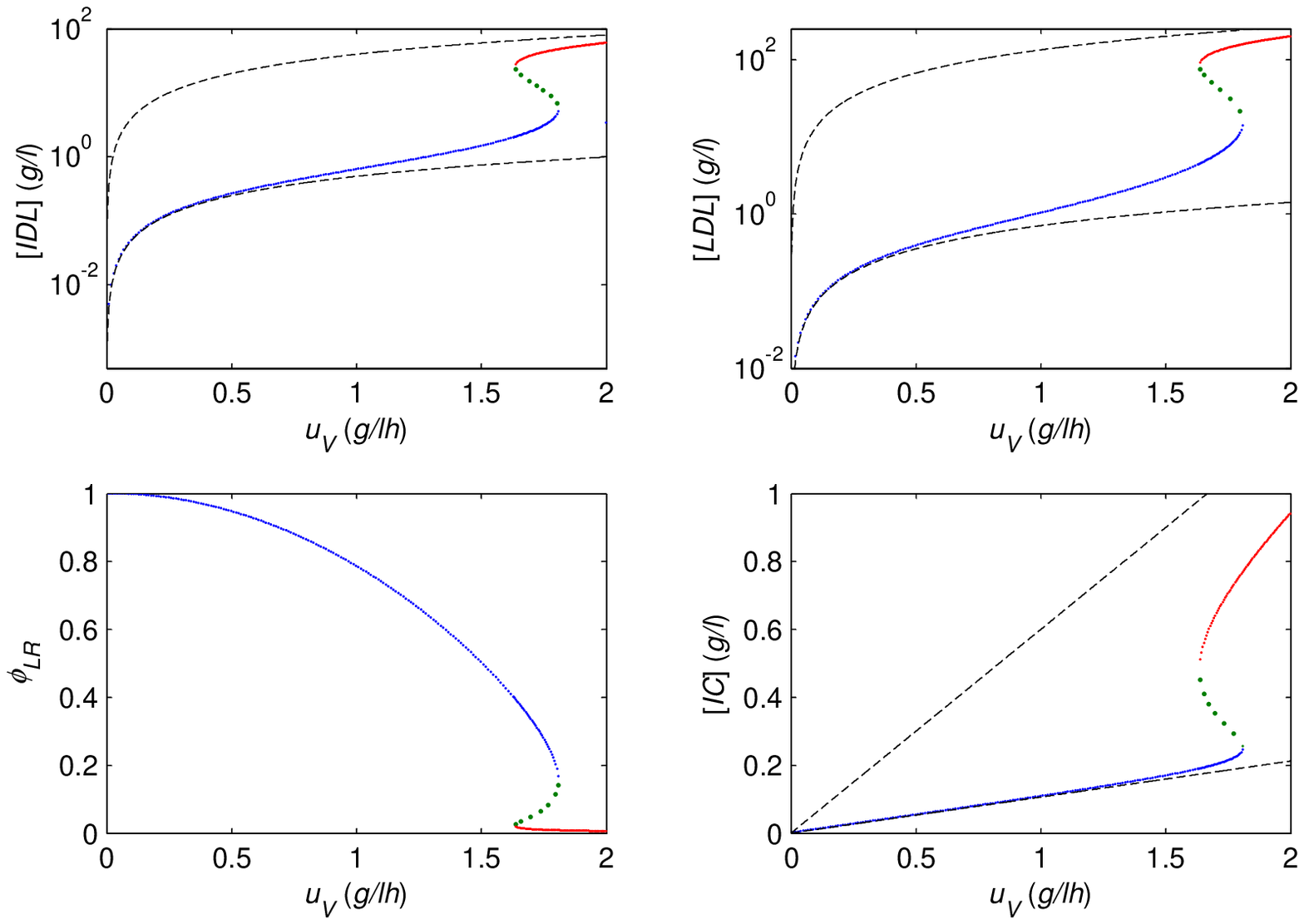}
\caption[\textbf{Plot of equilibrium solutions versus $u_V$.}]
{\textbf{Plot of equilibrium solutions versus $u_V$.} {\small We
show stable solutions (solid lines), unstable solutions (dotted
lines), and bounds~(\ref{bound1})--(\ref{bound4}) (dashed lines)
as a function of $u_V$ with parameters: \{ $k_V=0.3h^{-1}$,
$k_I=0.025h^{-1}$, $d_I=2h^{-1}$, $d_L=0.01h^{-1}$,
$d=0.0075h^{-1}$, $\chi_I=0.1$, $\chi_L=0.6$, $b=0.1lg^{-1}$,
$c=0.05g(lh)^{-1}$;  $d_{IC}=1h^{-1}$ \}. For this set of
parameters the system shows bistability. The bifurcations,
obtained through~(\protect{\ref{eq:quintic}}), occur at: $u_{V,
\text{bif}} \approx 1.637$ and $u_{V, \text{bif}} \approx 1.811$.
The approximation~(\protect{\ref{bif_reduced2}}) to the latter is
1.759.}} \label{bifd}
\end{figure}

The system is effectively four-dimensional since $[VLDL]$
converges exponentially to its equilibrium value
$[VLDL]_{eq}=u_V/k_V$.
Moreover, we can restrict our analysis to the non-negative orthant
because the system is essentially non-negative, i.e., solutions
remain non-negative for non-negative initial
conditions~\citep{WVE01}. The equilibrium point can also be
bounded from above using the nullclines
of~(\ref{full1})--(\ref{full5}) to give:
\begin{eqnarray}
\frac{u_V}{k_I+d_I}  \leq &[IDL]_{eq}&  \leq \frac{u_V}{k_I}\label{bound1}\\
\frac{k_I \, u_V}{(k_I+d_I) (d+d_L)} \leq &[LDL]_{eq}&  \leq \frac{u_V}{d}\label{bound2}\\
0  \leq &\phi_{LR_{eq}}&  \leq 1 \label{bound3} \\
\frac{\chi_Id_I+\chi_Lk_I}{k_I+d_I}\frac{u_V}{d_{IC}} \leq
&[IC]_{eq}&  \leq \frac{\chi_Lu_V}{d_{IC}} \label{bound4}.
\end{eqnarray}
The equilibrium approaches the
bounds~(\ref{bound1})--(\ref{bound4}) asymptotically as $u_V \to
0$ and $u_V \to \infty$, as shown in Figure~\ref{bifd}. The simple
linear bounds~(\ref{bound1})--(\ref{bound4}) represent well the
asymptotic behaviour although tighter nonlinear bounds (not shown)
can be obtained. In conclusion, there is a compact, connected,
positively invariant bounding region for the system. This implies
that any linearly stable solution in this region is also locally
asymptotically stable.

Consider now the fixed points of
system~(\ref{full1})--(\ref{full5}) which are the solutions to the
following algebraic equation:
\begin{small}
\begin{equation}  \label{eq:sol_full}
\begin{split}
&\phi_{LR_{eq}}^4d_{IC}cd_I^2d_L +\phi_{LR_{eq}}^3\left(d_{IC}cd_I(2k_Id_L+d_I(d-d_L))+u_V^2bd_I^2d_L\chi_I \right) \\
&+\phi_{LR_{eq}}^2 \left(d_{IC}c(2k_Id_I(d-d_L)+k_I^2d_L-d_I^2d)+u_V^2bd_I(d_Lk_I(\chi_I+\chi_L)+dd_I\chi_I)\right)\\
&+\phi_{LR_{eq}}\left(d_{IC}ck_I(k_I(d-d_L)-2d_Id)+u_V^2bk_I(dd_I\chi_L+d_Lk_I\chi_L)
\right) -d_{IC}ck_I^2d = 0
\end{split}
\end{equation}
\end{small} together with
\begin{small}
\begin{equation}
\begin{split} \left\{ [VLDL]_{eq} =\frac{u_V}{k_V}; \,\,
[IDL]_{eq}=\frac{u_V}{k_I+d_I\phi_{LR_{eq}}}; \,\,
[LDL]_{eq}=\frac{k_I[IDL]_{eq}}{d+d_L\phi_{LR_{eq}}}; \right.\\
\left.[IC]_{eq}=\frac{(\chi_Id_I[IDL]_{eq}+\chi_Ld_L[LDL]_{eq})\phi_{LR_{eq}}+\chi_Ld[LDL]_{eq}}{d_{IC}}
\right\} . \label{eq:sol_full2}
\end{split}
\end{equation}
\end{small}
Descartes' rule of signs implies that the number of positive
equilibria is 1 or 3. If we assume that the system has neither
(quasi)periodic solutions nor strange attractors (which never
appear in our numerics), then if there is only one fixed point, it
is globally asymptotically stable; and if there are three, two are
locally asymptotically stable and one is unstable and lies
``between'' the two stable points.

\begin{figure}[htbp]
\centering
\includegraphics[width=.95\textwidth]{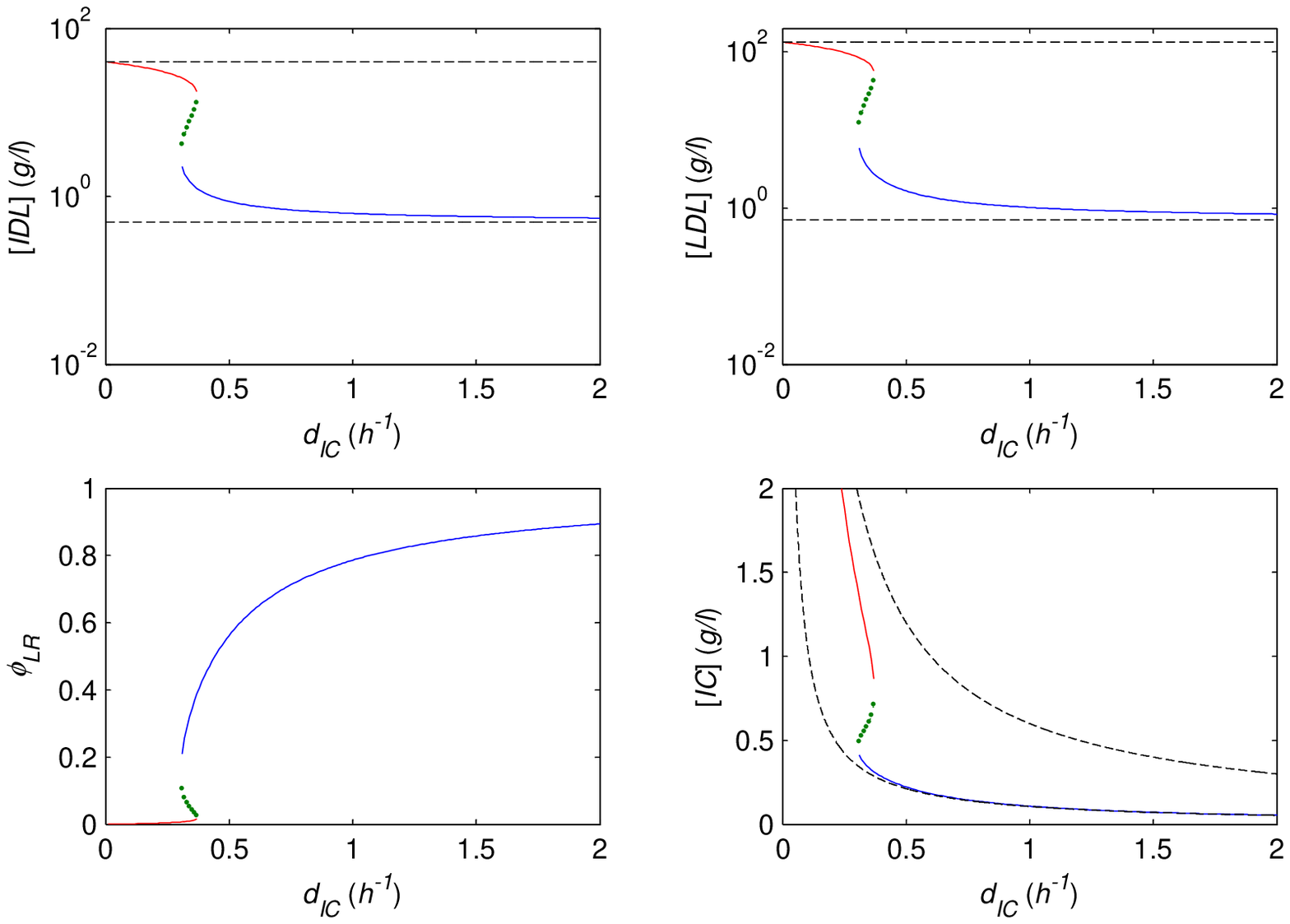}
\caption[\textbf{Equilibrium solutions versus $d_{IC}$.}]
{\textbf{Equilibrium solutions versus $d_{IC}$.} {\small We show
stable solutions (solid lines), unstable solutions (dotted lines),
and bounds (dashed lines) as a function of $d_{IC}$. All
parameters as in Fig.~\protect{\ref{bifd}} and $u_V=1
\,g(lh)^{-1}$. The bifurcations for this set of parameters,
obtained through~(\protect{\ref{eq:quintic}}), occur at: $d_{IC,
\text{bif}} \approx 0.373$ and $d_{IC, \text{bif}} \approx 0.305$.
The approximation~(\protect{\ref{bif_reduced2}}) to the latter is
0.323.}}\label{bifd2}
\end{figure}

We numerically explore the fixed point
equation~(\ref{eq:sol_full}) and record how the number and
stability of positive fixed points change as the control
parameters are varied. Figure~\ref{bifd} shows a numerical
bifurcation diagram as a function of $u_V$, for a given set of
parameters. The system has two distinct (sometimes coexisting)
stable solutions: a low and a high cholesterol branch. For small
(high) $u_V$, only the low (high) branch exists. For intermediate
values of $u_V$, the two stable branches (high and low) and an
unstable solution coexist. The existence of this coexistence
region leads to the possibility of hysteresis and a delayed return
to the normal state when $u_V$ is swept. The transitions that
delimit the coexistence region are two saddle-node bifurcations. A
similar bifurcation diagram (Figure~\ref{bifd2}), with equivalent
saddle-node bifurcations, is obtained by sweeping the other
control parameter $d_{IC}$, although in this case the transition
to a higher cholesterol state is produced by decreasing $d_{IC}$.
These bifurcations could be of physiological relevance because the
parameters $u_V$ and $d_{IC}$ can be regulated through diet and
medication.

The nature of the bifurcation diagram depends on the non-control
parameters of the model. In fact, the coexistence region does not
exist in certain regions of parameter space. In that case, the
transition from the low to the high cholesterol solution is
smooth, without undergoing a bifurcation (i.e., no sudden jumps
and no hysteresis). We make this more precise by analysing the
equation for the bifurcation points. Rewrite the fixed point
equation~(\ref{eq:sol_full}) as:
\begin{equation}
(y-1) (y+g) (y+f)^2 + Z \, y \, (y+h f) (y+f+g)  = 0
\label{eq:fullsystempolynomial}
\end{equation}
where $y={\phi_{LR}}_{eq}$ and the following ratios of parameters
have been defined:
\begin{equation} e=b/c, \, f=k_I/d_I, \, g = d/d_L, \,
h=\chi_L/\chi_I, \, Z=u_V^2\, e \, \chi_I/d_{IC}.
\label{eq:ratios}
\end{equation} The bifurcation points are the solutions of an augmented system of
equations, formed by the fixed point equation and the condition
that the Jacobian of~(\ref{full1})--(\ref{full5}) vanishes. Using
Groebner bases (or considerations about monotonicity and curvature
of~(\ref{eq:fullsystempolynomial})) we obtain a quintic polynomial
equation for the bifurcation points $y_{\text{bif}}$:
\begin{multline}
y_{\text{bif}}^5 + (2\,g + 2\,f\,h + f)\,y_{\text{bif}}^{4} +
(g^{2} + 3\,f^{2}\,h + f -
f\,h - g\,f + 4\,f\,h\,g)\,y_{\text{bif}}^{3} \\
\mbox{} + (4\,g\,f + f^{2} - f^{2}\,g + f^{3}\,h - 2\,f\,h\,g + 2
\,f^{2}\,h\,g - g^{2}\,f - f^{2}\,h + 2\,f\,h\,g^{2})\,y_{\text{bif}}^{2} \\
\mbox{} + (2\,f^{2}\,g - f\,h\,g^{2} + 2\,g^{2}\,f + f^{2}\,h\,g)
\,y_{\text{bif}} + f^{3}\,h\,g + f^{2}\,h\,g^{2}=0.
\label{eq:quintic}
\end{multline} Based on Descartes' rule, we conclude that there is
no coexistence region (and no bifurcations) if:
\begin{align*}
g^{2} + 3\,f^{2}\,h + f - f\,h - g\,f + 4\,f\,h\,g & > 0 \\
4\,g\,f + f^{2} - f^{2}\,g + f^{3}\,h - 2\,f\,h\,g + 2
\,f^{2}\,h\,g - g^{2}\,f - f^{2}\,h + 2\,f\,h\,g^{2} & > 0 \\
2\,f^{2}\,g - f\,h\,g^{2} + 2\,g^{2}\,f + f^{2}\,h\,g & >  0.
\end{align*}
These inequalities can be tested for each set of parameters to
establish a sufficient condition for the saddle-node bifurcations
to exist.

Further insight can be gathered through explicit
\textit{approximate} formulas for the bifurcations. Consider the
limit case with $d=0$, i.e., when there is no direct uptake of
cholesterol by the cell and only receptor-mediated intake is
possible. In this case, the fixed point
equation~(\ref{eq:fullsystempolynomial}) simplifies to:
\begin{equation*}
y \, (y+f) \, \left(y^2+(f-1+Z) y +f (Z h -1)\right) = 0,
\end{equation*}
leading to the following bifurcation points for this limit case:
{\small \begin{align}  4f(Z h-1)& = 0 &\Rightarrow
u_{V, \text{bif}}^*&= \sqrt{\frac{d_{IC}}{e \chi_I \, h}} \label{bif_reduced1} \\
\frac{(f-1+Z)^2}{f (Z h -1)} & = 4  &\Rightarrow u_{V,
\text{bif}}^*&= \sqrt{\frac{d_{IC}}{e \chi_I} \left(1-f+2 f h - 2
\sqrt{f (h-1)(f h+1)}\right)}. \label{bif_reduced2}
\end{align}}
Eq.~(\ref{bif_reduced2}) marks the point above which the low
cholesterol branch ceases to exist and serves as a good
approximation for the bifurcation of the full model if $d$ is
small. The approximate
formulas~(\ref{bif_reduced1})--(\ref{bif_reduced2}) explain the
overall dependence on $u_V$ and $d_{IC}$ of the bifurcation points
shown in Figs.~\ref{bifd}~and~\ref{bifd2}.

\subsection{Sensitivity Analysis of the Equilibrium Solution}\label{sens}

In order to ascertain how variations in parameter values (e.g.,
among individuals or due to age) affect the results, we perform a
sensitivity analysis of the system. Because of its physiological
relevance, we focus on the variability of the low cholesterol
branch under parameter uncertainty.

{\small \begin{table} \centering
\begin{tabular}{|c|c|c|c|}

\hline \multirow{2}{*}{Parameter} & \multirow{2}{*}{Variable} &
\multicolumn{2}{c|}{ Sensitivity over lower branch ($u_V \leq
u_{V, \text{max}}$)}\\ \cline{3-4} & & Median & Range \\
\hline
\vspace{-.2cm} $k_I\in[0.025,0.1]$ & $[IDL]$ & 0.027 & $[0.0038, 0.036]$ \\
\vspace{-.2cm} ($u_{V, \text{max}} =1.1$)& $[LDL]$ & 1.18 & $[1.16, 1.26]$ \\
\vspace{-.2cm}  & $\phi_{LR}$ & 0.013 & $[3.1\times10^{-4}, 0.074]$ \\
& $[IC]$ & 0.17 & $[0.15, 0.21]$ \\
\hline
\vspace{-.2cm} $d_I\in[0.5,2]$ & $[IDL]$ & 1.83 & $[1.79, 2.23]$  \\
\vspace{-.2cm} ($u_{V, \text{max}} = 1.4$)& $[LDL]$ & 1.85 & $[1.79, 2.48]$  \\
\vspace{-.2cm}  & $\phi_{LR}$ & 0.023  & $[3.1\times10^{-4}, 0.22]$ \\
& $[IC]$ & 0.18 & $[0.16, 0.34]$ \\
\hline
\vspace{-.2cm} $d_L\in[0.005,0.02]$ & $[IDL]$ & $5.3\times10^{-4}$ & $[8.5\times10^{-6}, 0.0038]$  \\
\vspace{-.2cm} ($u_{V, \text{max}} = 1.3$)& $[LDL]$ & $0.82$ & $[0.67,0.87]$ \\
\vspace{-.2cm}  & $\phi_{LR}$ & $5.4\times10^{-4}$ & $[8.6\times10^{-6}, 0.0039]$ \\
 & $[IC]$ & $3.4\times10^{-5}$ & $[5.0\times10^{-7}, 3.4\times10^{-4}]$ \\
\hline
\vspace{-.2cm} $d\in[0.0025,0.0075]$ & $[IDL]$ & $4.6\times10^{-4}$ & $[5.9\times10^{-6},0.0045]$ \\
\vspace{-.2cm}  ($u_{V, \text{max}} = 1.4$)& $[LDL]$ & 0.38 & $[0.34, 0.50$] \\
\vspace{-.2cm}  & $\phi_{LR}$ & $4.7\times10^{-4}$ & $[6.0\times10^{-6},0.0046]$ \\
 & $[IC]$ & $3.0\times10^{-5}$ & $[3.5\times10^{-7},4.3\times10^{-4}]$\\
\hline
\vspace{-.2cm} $e = b/c \in[0.25,2.5]$ & $[IDL]$ & 0.15 & $[0.0024, 0.80]$  \\
\vspace{-.2cm} ($u_{V, \text{max}} = 1.4 $)& $[LDL]$ & 0.23 & $[0.0037, 1.23]$ \\
\vspace{-.2cm}  & $\phi_{LR}$ & 0.15 & $[0.0024, 0.70]$ \\
 & $[IC]$ & 0.0093 & $[1.4\times10^{-4}, 0.064]$ \\
\hline
\vspace{-.2cm} $\chi_I\in[0.25,0.45]$ & $[IDL]$ & 0.12 & $[0.0039, 0.76]$  \\
\vspace{-.2cm} ($u_{V, \text{max}} = 0.9$)& $[LDL]$ & 0.18 & $[0.0062, 1.06]$\\
\vspace{-.2cm}  & $\phi_{LR}$ & 0.12 & $[0.004, 0.71]$ \\
 & $[IC]$ & 0.56 & $[0.56, 0.56]$ \\
\hline
\vspace{-.2cm} $\chi_L\in[0.4,0.5]$  & $[IDL]$ & 0.0018 & $[2.4\times10^{-5}, 0.014]$  \\
\vspace{-.2cm} ($u_{V, \text{max}} = 1.4$)& $[LDL]$ & $0.0027$ & $[3.8\times10^{-5}, 0.21]$ \\
\vspace{-.2cm}  & $\phi_{LR}$ & $0.0018$ & $[2.5\times10^{-5}, 0.015]$ \\
& $[IC]$ & $0.013$ & $[0.012, 0.021]$ \\
\hline
\end{tabular}\\ \caption{\textbf{Sensitivity of the variables of the model
on the lower branch to variation of parameters one at a time.}
{\small The sensitivity is calculated as the range divided by the
median of each variable for five different values of the parameter
that appears in the first column. The third and fourth columns
report the median sensitivity and the range of calculated
sensitivities over the lower branch (up to the value $u_{V,
\text{max}}$). The salient feature of the analysis is the high
sensitivity of $[LDL]$ to all parameters and the relative
insensitivity of $[IC]$ to almost all parameters.
}} \label{Table3}
\end{table}}

Table~\ref{Table3} summarises the sensitivity of the low
cholesterol branch when parameters are modified one at a time.
Consider for instance the effect of $d$, which parameterises the
non-specific uptake of LDL. We have calculated numerical
bifurcation diagrams similar to those in Figure~\ref{bifd} for
values of $d \in [0.0025,0.0075]$. As $d$ is varied, the values of
the lower and upper branches and the location of the bifurcations
change. However, as expected from~(\ref{bound1})--(\ref{bound4}),
only the $[LDL]$ is affected significantly. To make this more
precise, we calculate the sensitivity for each variable over the
lower branch ($u_V \leq 1.4 \,g(lh)^{-1}$) by obtaining the
average ratio of the range and median of the equilibrium values.
Clearly, $[LDL]$ has the highest sensitivity to $d$.  The
sensitivity to other parameters is summarised in
Table~\ref{Table3}: $d_L$ virtually only affects $[LDL]$, as
indicated by~(\ref{bound1})--(\ref{bound4}); variations of $e=b/c$
displace the bifurcation point~(\ref{bif_reduced2}), but $[IC]$ is
fairly insensitive; variations in $k_I$ and $d_I$ have a large
effect on $[LDL]$, but only $d_I$ affects $[IDL]$; the sensitivity
of $[IC]$ to $\chi_I$ is moderate as compared to the sensitivity
of $[LDL]$ or $[IDL]$ to $k_I$ and $d_I$.

We have also performed a Monte Carlo sensitivity analysis with
respect to variations of all parameters at once by sampling and
averaging over the uniform hypercube of non-control parameters in
Table~\ref{Table3}. The results show that the median sensitivity
of $[LDL]$ to global parametric variations is an order of
magnitude larger than that of $[IC]$ over the low cholesterol
branch. The conclusion of the sensitivity analysis is that while
the plasma cholesterol $[LDL]$ is sensitive to most parameters,
the intracellular cholesterol $[IC]$ is robust to parameter
perturbations. In essence, this is a system where $[IC]$ is
tightly controlled whereas $[LDL]$ (the variable measured
medically) is poorly controlled and can undergo wide variations.

Based on experimental observations~\citep{JM77}, we have modelled
the regulation of the nuclear synthesis of LDL receptors as a
simple reciprocal feedback. However, the qualitative features of
the system are robust to our choice of functional form. The
regulation term in Eq.~(\ref{full4}) is a particular case of the
standard inhibiting feedback~\citep{MU90}: $c
(1-\phi_{LR})/(q+[IC]^n)$, with $n=1$ and $q \ll [IC]$. Note,
however, that the presence of $q$ and $n$ does not change the
bounds~(\ref{bound1})--(\ref{bound4}). Therefore, the low and high
cholesterol asymptotic regimes and the asymptotic response of the
system to time-varying inputs
(Section~\ref{section:time-dependent}) remain unchanged. We have
also checked numerically that the system still undergoes
transitions between low and high cholesterol states; $q$ and $n$
have the effect of displacing the bifurcation points and modifying
the extent of the coexistence region. In general, for large $q$
the effect of the feedback is reduced, thus making the transition
to the high cholesterol branch easier; for $n > 1$ the feedback is
more efficient, thus enlarging the region of existence of the
lower cholesterol branch.

\section{Physiological Implications of the Model}\label{section:physio}
\subsection{Parametric Effects of Medical and Genetic Conditions}
\subsubsection{Tangier Disease and Reverse Cholesterol Transport}

\emph{Tangier disease} is a deadly disease in which vanishing
reverse cholesterol transport leads to accumulation of cholesterol
in all cells~\citep{SGYCJF99}. In our model, the parameter
$d_{IC}$ depends directly on reverse cholesterol transport. Note
that only the bounds on $[IC]_{eq}$ are inversely proportional to
$d_{IC}$. This means that low values of $d_{IC}$ lead to diverging
levels of intracellular cholesterol, while the plasma cholesterol
levels saturate (Fig.~\ref{bifd2}).

\subsubsection{Lipase Enzyme Activity}

In a review article, \cite{YSOS03} report that reduced enzyme
activity can lead to reduced plasma cholesterol concentrations in
mice and rabbits. Similar results might hold for humans. A
conclusion of our sensitivity analysis (Table~\ref{Table3}) is
that variations in $k_I$ produce sharp changes in $[LDL]$ without
affecting the other variables significantly. Further numerics (not
shown) show that decreasing $k_I$ decreases $[LDL]$.

\subsubsection{Familial Hypercholesterolemia}\label{FH}

{\it Familial hypercholesterolemia} (FH) is a genetic disease
caused by mutations in the LDL receptor gene~\citep{JM77,HHH90}.
In the heterozygous case of FH, which affects about 1 in 500
people in generic populations and even higher ratios in more
inclusive populations, cells express around half the normal number
of functional receptors on their surface leading to a two-fold
increase in $[LDL]$. The homozygous case is much more rare (about
1 per million) but more severe. In this case, a negligible amount
of functional LDL receptors lead to a dramatic rise of LDL plasma
levels. These diseased individuals frequently dye of myocardial
infarction or stroke in their teens.  Table~\ref{Table4} presents
a classification of FH gene mutations with their physiological
implications~\citep{HHH90}. We also show how the mutations can be
represented in terms of the parameters of our model.
\begin{table}[htbp]
\centering
\begin{tabular}{|c|l|c|}
\hline
Mutation Class& Physiological effect & Affected Parameter\\
\hline
1 & Reduced synthesis of LR&\multirow{2}{*}{$c\downarrow$}\\
2 & Reduced transport of LR & \\
\hline 3 & Reduced binding of IDL/LDL to LR &
\multirow{2}{*}{$d_I\downarrow$ and/or $d_L\downarrow$} \\
4 & Reduced number of LR in coated pits & \\
\hline
5 & Diminished recycling of LR & $b\uparrow$\\
\hline
\end{tabular}\\ \caption{\textbf{Gene mutations that lead
to familial hypercholesterolemia.} \small Mutations leading to FH
as classified by~\cite{HHH90} and how they affect the parameters
of the model.} \label{Table4}
\end{table}

Most of the medical observations in~\cite{HHH90} refer to mutation
classes 3 and 4, both hetero- and homozygous. In terms of our
model, heterozygous mutations 3 and 4 would halve $d_I$ and $d_L$.
This produces roughly a two-fold (up to four-fold) increase in
$[LDL]$, as deduced from the lower bound in~(\ref{bound2}) and
Table~\ref{Table1}. One of the homozygous mutations blocks the
binding of both IDL and LDL to LR, hence $d_I=d_L=0$. Again
from~(\ref{bound2}), this would increase $[LDL]$ by a factor
$(1+d_L/d)(1+d_I/k_I)$, which is roughly 10-fold (and up to
several hundred times-fold). These numerical factors correspond
well to those in the literature~\citep{HHH90}. Under another
homozygous mutation 3 and 4, LDL receptors bind to IDL but not to
LDL, i.e., $d_L=0$. This produces an increase of $[LDL]$ in the
lower branch by a factor of $(1+d_L/d)$, which is two-fold (up to
nine-fold) but still an order of magnitude smaller than that of
the other homozygous case. Moreover, the characteristic time scale
for the catabolism of LDL (see
Section~\ref{section:time-dependent}) increases from
$(d+d_L)^{-1}$ (around two days) to $d^{-1}$ (over a week). Again,
this coincides with experimental observations that the removal of
LDL from the plasma is delayed and the increase in plasma LDL is
moderate~\citep{HHH90}.

Medical evidence for mutations 1-2 and 5 is scarcely reported. In
our model, they would translate into an increase of $e=b/c$ which
would have a small effect on the plasma concentrations on the
lower cholesterol branch. However, as seen in
Eqs.~(\ref{bif_reduced1})~and~(\ref{bif_reduced2}), an increase in
$e$ would lower the bifurcation point, making it easier to jump to
the high cholesterol branch in response to an increase in $u_V$ or
a decrease in $d_{IC}$. Therefore, under these mutations the
parameter region where the lower cholesterol regime exists would
be reduced.

Other observations in~\cite{HHH90} could also be interpreted in
terms of our model. For example, it was reported that despite
treatment with statins and lipid-lowering medications, the plasma
cholesterol of some patients remained between 600 and $1200
\,mg/dl$, five to ten times the normal levels. As seen in
Figs.~\ref{bifd}~and~\ref{bifd2} and discussed further in
Section~\ref{section:slow_drift}, the presence of bistability
means that once on the upper branch it is more difficult to return
to the low cholesterol state by reducing $u_V$, e.g.~through
statin medication. Another intriguing observation is the dramatic
clinical variability among FH patients with the same genetic
mutation. A possible explanation is the robustness of $[IC]_{eq}$
to variations in the physiological parameters. Indeed, from the
lower bound in~(\ref{bound4}) we estimate that $[IC]$ would only
increase by 1.1 to 1.5 times under homozygotic FH, while the
corresponding increase in $[LDL]$ would range from one to almost
three orders of magnitude. This would allow the development of
individuals who would have similar levels of intracellular
cholesterol while the LDL plasma levels could be very high in some
cases, with different clinical consequences.

\subsection{Response to Time-dependent Inputs}
\label{section:time-dependent}

The time scales of the system span from a few hours up to a week.
These time scales govern the way in which the system responds to
external drives or perturbations to the variables. To clarify
their dynamical role, we use singular perturbation methods to
obtain slow and fast manifolds~\citep{HKK00} relevant to the
system. We have seen previously that the lower bound~($\phi_{LR} =
1$) and upper bound~($\phi_{LR} = 0$) are approached
asymptotically as $u_V \to 0$ and as $u_V \to \infty$,
respectively. In these singular asymptotic limits, the
system~(\ref{full1})--(\ref{full5}) reduces to a linear system of
the form:
\begin{equation} \frac{d\mathbf{x}}{dt}=A_k \,\mathbf{x}+ \mathbf{b},
\quad k=\{l,u\}, \label{eq:linearasymptotic}
\end{equation}
where $\mathbf{x}=[[VLDL]\ [IDL]\ [LDL]\ [IC]]^\mathrm{T}, \quad
\mathbf{b}=[u_V\ 0\ 0\ 0]^\mathrm{T}$,
\begin{displaymath}
A_l=\left[ \begin{array}{cccc}
-k_V & 0 & 0 & 0 \\
 k_V & -k_I-d_I & 0  & 0 \\
   0 &  k_I & -d_L-d & 0 \\
   0 & \chi_Id_I & \chi_L(d_L+d) & -d_{IC} \\
\end{array} \right], \, \ A_u=\left[ \begin{array}{cccc}
-k_V & 0 & 0 & 0 \\
 k_V & -k_I & 0  & 0 \\
   0 &  k_I & -d & 0 \\
   0 & 0 & \chi_Ld & -d_{IC} \\
\end{array} \right],
\end{displaymath}
and $l, u$ denote \textit{lower} and \textit{upper} asymptotic
regimes, respectively. The inverse of the matrix eigenvalues
are the characteristic times for each variable to return to
equilibrium:
\begin{multline} \label{eq:timescales}
    \left(k_I+d_I\right)^{-1}\approx 0.5  <  d_{IC}^{-1}
    \approx
    1\, h < k_V^{-1} \approx 10 \\
  < k_I^{-1}\approx 40 <  \left(d+d_L\right)^{-1}\approx 70  <
  d^{-1}\approx
  200,
\end{multline}
all of them given in hours. Note that the slowest time scale
increases from $(d+d_L)^{-1}$ (on the scale of a day) on the lower
branch to $d^{-1}$ (over one week) on the upper branch.

\subsubsection{Response to slow drifts in the control parameters}
\label{section:slow_drift}

\begin{figure}
\centering
\includegraphics[width=.85\textwidth]{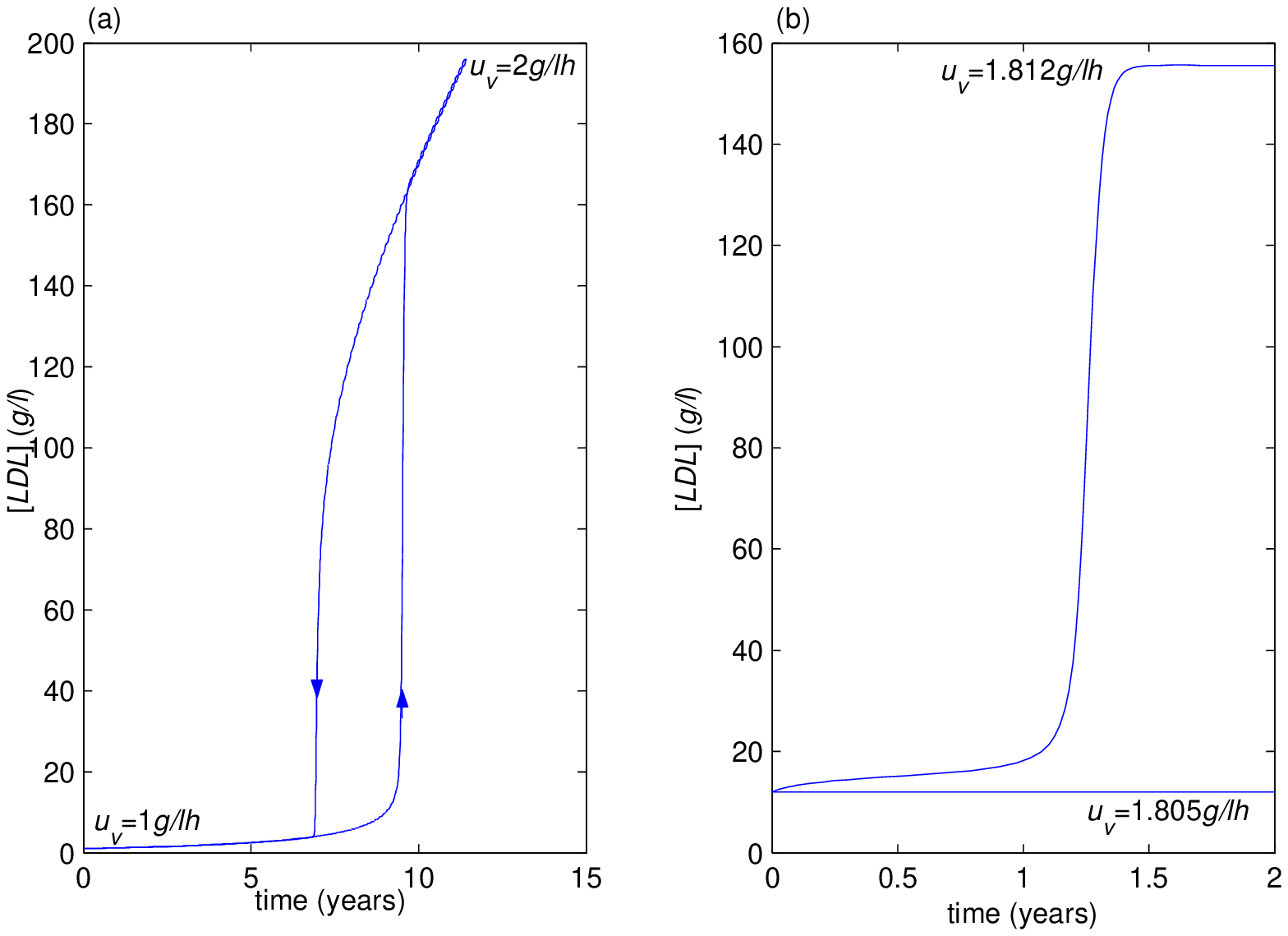}
\caption{{\small \textbf{Quasi-static nonlinear effects} (a)
Hysteresis curve obtained when $u_V$ is increased by $0.01 \,
g(lh)^{-1}$ every 1000 hours from $u_V=1 \, g(lh)^{-1}$ to $u_V=2
\, g(lh)^{-1}$, and then decreased at the same rate to its initial
value. (b) Bottleneck dynamics caused by the `ghost' saddle-node
around the bifurcation. The jump to the upper branch occurs a year
after a small, but permanent, perturbation of $u_V$ has occurred.
}}\label{fig:Hyster_Ghost}
\end{figure}

The response of the system to a slowly drifting input $u_V(t)$,
which varies on a time scale slower than $d^{-1}$, will be
quasi-static, tracking the equilibrium points. This can reveal
nonlinear behaviour, such as hysteresis and bottleneck
phenomena~\citep{SHS94}, which can have physiological
ramifications.

If the system is in a parametric region where it can undergo
bifurcations (as in Figs.~\ref{bifd} and \ref{bifd2}), it will
exhibit standard hysteresis when $u_V$ is swept continuously from
the lower branch to the upper branch and backwards
(Figure~\ref{fig:Hyster_Ghost}a). This means that once in the
upper branch it is more difficult to return to the low cholesterol
state by reducing $u_V$ (e.g., through statin medication).

Another effect of the existence of saddle-node bifurcations is the
possibility of bottleneck dynamics caused by saddle-node ghosts.
This can translate into long-lived transients when the system is
parametrically close to the bifurcation points.
Figure~\ref{fig:Hyster_Ghost}b shows the time evolution of the
system for two values of $u_V$, one below and one above (but very
close) to the bifurcation. The former remains in the low
cholesterol branch and the latter jumps the high cholesterol
solution. However, although the actual transition is relatively
sudden (on the order of a week, $d^{-1}$), it occurs after a
prolonged period of more than a year in which the system behaves
as if in the (no longer existing) low cholesterol state. This
pseudo-steady (ghost) behaviour would make it difficult to detect
a very slow $u_V$ drift, which can suddenly lead to high plasma
cholesterol values, if clinical measurements are infrequent.

\subsubsection{Response to periodic inputs}
\begin{figure}[h]
\begin{picture}(0,280)
\includegraphics[width=1\textwidth]{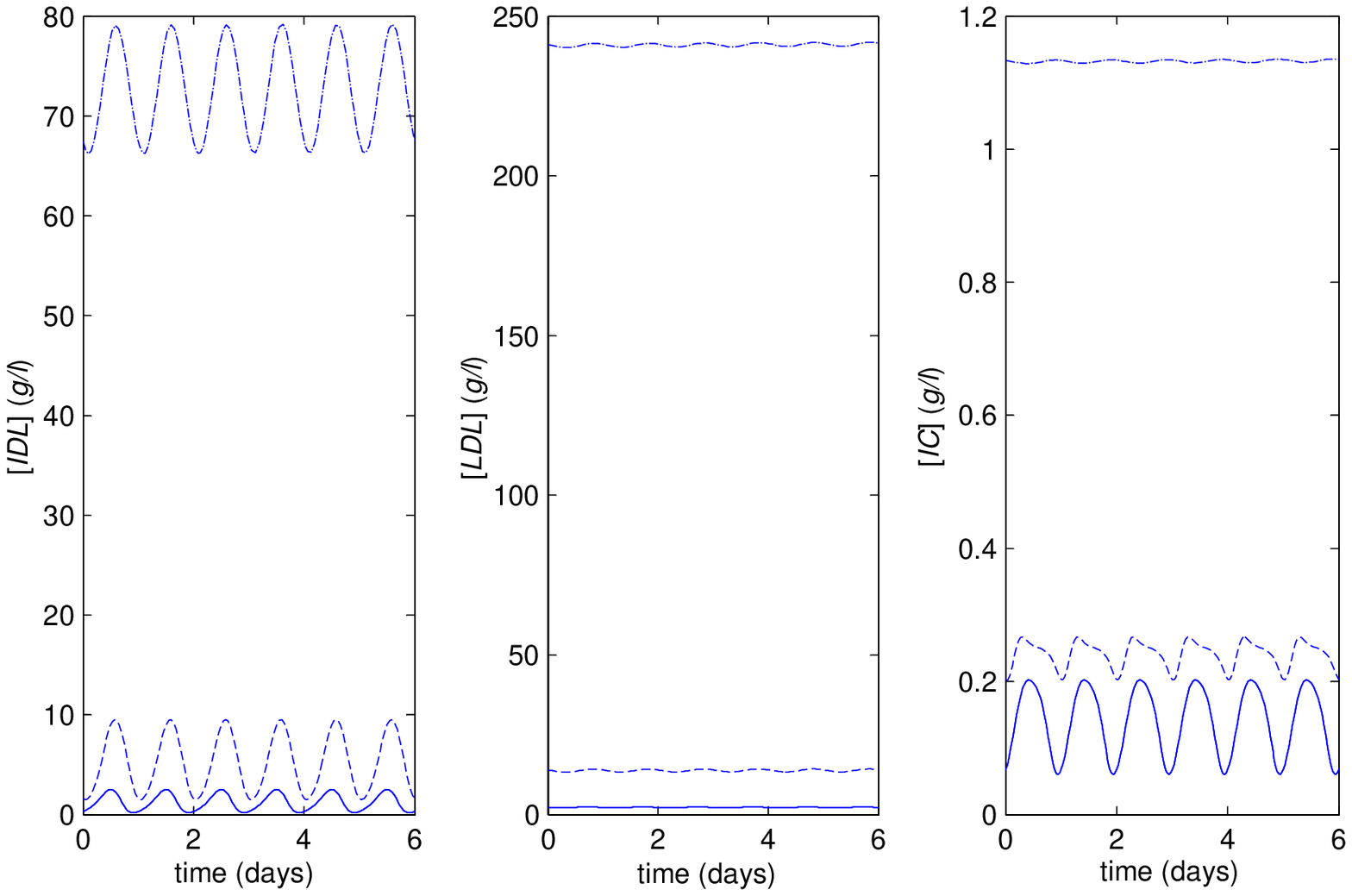}
\end{picture}
\caption[\textbf{Response to forced oscillations with a period of
one day.}] {\textbf{Forced oscillations with a period of one day.}
\small Response of $[IDL]$, $[LDL]$ and $[IC]$ to a periodic
forcing $u_V=U_V \, \left (1+\sin(\omega \, t)\right )$ with
$\omega = 2 \pi/24 \, h^{-1}$ for values of $U_V=$ 1.25 (solid
line), 1.75 (dashed line), 2.25 (dot-dashed line). All other
parameters as in Fig.~\protect{\ref{bifd}. The variables of the
system exhibit different dynamical responses at this forcing
frequency. The variability of $[IC]$ is linked to that of $[IDL]$
on the lower branch and to that of $[LDL]$ on the upper branch.}
}\label{graph6}
\end{figure}

We study now the physiological implications of the response of the
system to forced input oscillations with periods of days up to
months. It is reasonable to assume that the production of VLDL
($u_V(t)$) can have a strong periodic content on the order of a
day since VLDL is produced at increased levels during the night;
dietary intake is often highly cyclic; and the ingestion of
statins to lower $u_V$ generally follows a daily cycle. As
expected, entrainment phenomena can be observed in
Figure~\ref{graph6}, where we show the response of the system to
forced oscillations with period of one day and increasing
amplitude and mean. $[LDL]$ is relatively insensitive to the
oscillatory drive, while $[IDL]$ exhibits large oscillatory
amplitudes both for small and large drives. However, the salient
feature is the different amplitude of the response of $[IC]$ in
both regimes. Qualitatively, this can be understood from the fact
that $[IC]$ changes from being tightly linked to $[IDL]$ at low
drives, when the amount of LDL receptors is high, to being
governed by $[LDL]$ on the upper branch, where the low $\phi_{LR}$
makes IDL endocytosis virtually impossible while LDL can still be
absorbed through the non-receptor mediated route.

A quantitative understanding of the responses in
Figure~\ref{graph6} 
emerges from a more detailed analysis. If we consider a periodic
input
\begin{equation}
u_V(t)=U_V \left(1+P \sin(\omega\, t)\right),
\label{eq:forcing}
\end{equation}
it follows that the time response of the asymptotic
system~(\ref{eq:linearasymptotic}) is sinusoidal of the form
\begin{equation} \label{eq:periodic_asymptotic}
x_i=x_{i,av} +x_{i,s} \sin(\omega \, t)+ x_{i,c} \cos(\omega \,
t), \qquad i=1,\ldots,4,
\end{equation}
with average $x_{i,av}$ and amplitude
$x_{i,amp}=\sqrt{x_{i,s}^2+x_{i,c}^2}$. We have calculated the
sinusoidal responses for $[IDL], [LDL]$ and $[IC]$ on both
branches. On the lower branch we get:
\begin{small}
\begin{align}
[IDL]_{av,l}&=\frac{U_V}{k_I+d_I}, \quad
\frac{[IDL]_{amp,l}}{[IDL]_{av,l}} = \frac{P}{\sqrt{
\left(1+\left(\frac{\omega}{k_V}\right)^2\right)
\, \left(1+\left(\frac{\omega}{k_I+d_I}\right)^2\right)}} \label{eq:per_asymp_low1}\\
[LDL]_{av,l}&=\frac{k_I \, U_V}{\left(k_I+d_I\right)
\left(d+d_L\right)} , \quad \frac{[LDL]_{amp,l}}{[LDL]_{av,l}} =
\frac{[IDL]_{amp,l}}{[IDL]_{av,l}} \frac{1}{\sqrt{1+\left(\frac{\omega}{d+d_L}\right)^2}}  \label{eq:per_asymp_low2}\\
[IC]_{av,l}&=\frac{(\chi_I d_I+\chi_L k_I) \,
U_V}{\left(k_I+ d_I \right) d_{IC}}, \nonumber \\
 \quad \quad \quad & \frac{[IC]_{amp,l}}{[IC]_{av,l}} =
\frac{[IDL]_{amp,l}}{[IDL]_{av,l}} \sqrt{\frac{1+\left(\frac{
\omega}{d+d_L} \,\frac{\chi_I d_I}{\chi_I d_I+\chi_L
k_I}\right)^2}{\left(1+\left(\frac{\omega}{d+d_L}\right)^2\right)
\, \left(1+\left(\frac{\omega}{d_{IC}}\right)^2\right)}},
\label{eq:per_asymp_low3}
\end{align}
\end{small}
while on the upper branch we obtain:
\begin{align}
[IDL]_{av,u}&=\frac{U_V}{k_I}, &
\frac{[IDL]_{amp,u}}{[IDL]_{av,u}} &= \frac{P}{\sqrt{
\left(1+\left(\frac{\omega}{k_V}\right)^2\right)
\left(1+\left(\frac{\omega}{k_I}\right)^2\right)}}  \label{eq:per_asymp_up1}\\
[LDL]_{av,u}&=\frac{U_V}{d} , & \frac{[LDL]_{amp,u}}{[LDL]_{av,u}}
&= \frac{[IDL]_{amp,u}}{[IDL]_{av,u}} \frac{1}{\sqrt{1+\left(\frac{\omega}{d}\right)^2}} \label{eq:per_asymp_up2} \\
[IC]_{av,u}&=\frac{\chi_L U_V}{d_{IC}}, &
\frac{[IC]_{amp,u}}{[IC]_{av,u}} &=
\frac{[LDL]_{amp,u}}{[LDL]_{av,u}}
\frac{1}{\sqrt{1+\left(\frac{\omega}{d_{IC}}\right)^2}}
\label{eq:per_asymp_up3}.
\end{align}
The sinusoidal
solutions~(\ref{eq:periodic_asymptotic})--(\ref{eq:per_asymp_up3}),
which are valid asymptotically in the limits $u_V \to 0$ (lower)
and  $u_V \to \infty$ (upper), provide a good approximation when
$U_V$ is either small or large. Clearly, one does not expect a
pure sinusoidal response at intermediate values of $U_V$, as can
be observed in Figure~\ref{graph6}.

The change in the variability of $[IC]$ on both branches can now
be traced to
Eqs.~(\ref{eq:per_asymp_low1})--(\ref{eq:per_asymp_up3}). For a
forcing period of one day ($\omega = \pi/12 \approx 0.26 \,
h^{-1}$) and using the time scales~(\ref{eq:timescales}), the key
terms are $\omega/(d+d_L)$ and $\omega/d$. Note how on the lower
branch~(\ref{eq:per_asymp_low3}), the $\omega/(d+d_L)$ terms
cancel, with the effect that the relative variability
$[IC]_{amp,l}/[IC]_{av,l}$ is tied to that of $[IDL]$. Onn the
upper branch~(\ref{eq:per_asymp_up3}), $\omega/d_{IC} \ll 1$,
which implies that the relative variability
$[IC]_{amp,u}/[IC]_{av,u}$ is linked to that of $[LDL]$.

\begin{figure}[h]
\centering
\includegraphics[width=0.85\textwidth]{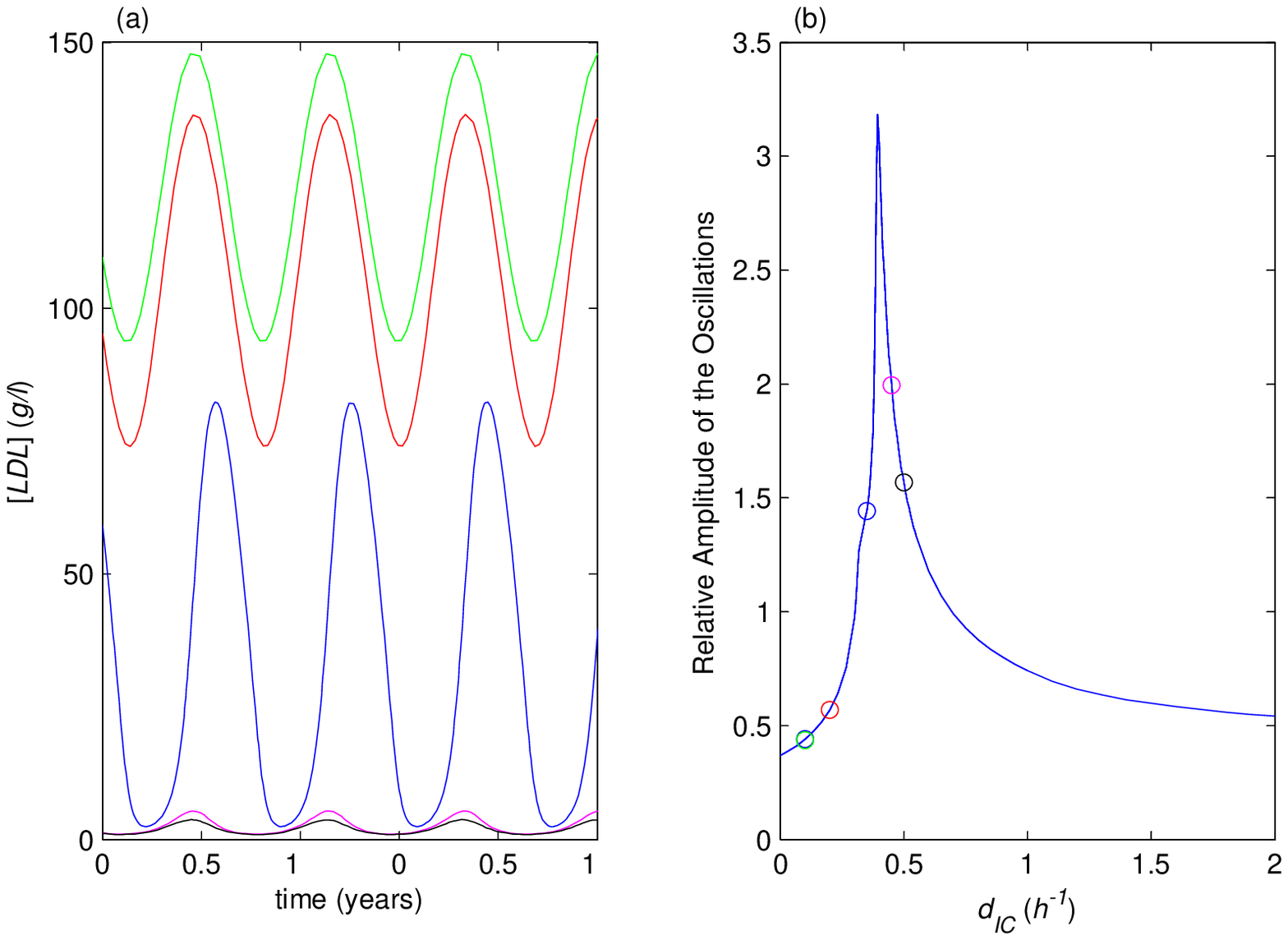}
\caption {\textbf{Forced oscillations with a period of 100 days.}
\small (a) Response of $[LDL]$ to a periodic forcing $u_V=\left(
1+0.2\,\sin(\omega \, t) \right )$ with $\omega = 2 \pi/2400 \,
h^{-1}$ for values of $d_{IC}=[0.1, 0.2, 0.35, 0.45, 0.5]$, from
top to bottom. All other parameters as in
Fig.~\protect{\ref{bifd2}}. As $d_{IC}$ increases, the system
moves from the upper to the lower branch and the average level of
$[LDL]$ decreases. (b) Relative amplitude of the oscillations,
calculated as the ratio of range and mean of the oscillations from
numerical simulations. The circles correspond to the values of
$d_{IC}$ in (a). Note that the relative amplitude is similar in
both asymptotic regimes. This can be understood from
Eqs.~(\protect{\ref{eq:per_asymp_low2}})~and~(\protect{\ref{eq:per_asymp_up2}}),
whence we obtain the relative amplitude to be $\approx 0.38$ (as
$d_{IC} \to 0$) and $\approx 0.40$ (as $d_{IC} \to \infty$). The
relative amplitude reaches a maximum close to the bifurcation,
where the system jumps between the upper and lower branches and is
at its most nonlinear. As an indication, the bifurcation for these
parameters and $u_V=1$ occurs at $d_{IC}\approx 0.373$, close to
the observed peak.}\label{100days}
\end{figure}

These observations suggest that the correlation between
intracellular and plasma cholesterol concentrations could be used
as a marker to detect the high cholesterol state. This effect
would be most pronounced in hepatic cells but might still be
observable in other cells and detectable through a skin test.
(This would depend on how much IDL can penetrate peripheral cells
by leaking through the capillary endothelium into the interstitial
fluid.)  A less practical alternative would imply repeated time
measurements of $[IDL]$ to check for large amplitude oscillations.

We have also explored the response of the system to oscillations
with lower frequency, as would be induced by seasonal cycles.
Figure~\ref{100days} shows the response of $[LDL]$ to oscillatory
inputs with period of 100 days at different values of $d_{IC}$ as
shown in Fig.~\ref{bifd2}. Consistent with
Eqs.~(\ref{eq:per_asymp_low2})~and~(\ref{eq:per_asymp_up2}), the
average level of $[LDL]$ on the upper branch is high, of order
$\sim d^{-1} \approx 133$, while on the lower branch it is much
smaller, of order $\sim k_I d_I^{-1} (d+d_L)^{-1} \approx 0.7$.
However, the \emph{relative} amplitude of the oscillations of
$[LDL]$ in both asymptotic regimes is of the same order. It is
precisely at the bifurcation point, when the system oscillates
between the low and high cholesterol regimes, that the
oscillations are larger in relative terms (Figure~\ref{100days}).
This could be used to characterise the system if sufficiently
frequent measurements are available.

\section{Discussion}\label{section:conclusions}

Understanding lipoprotein metabolism is important because high
lipoprotein concentration in the plasma is a major risk factor for
atherosclerosis, the most common cause of death in Western
societies. We have introduced a dynamical model of lipoprotein
metabolism derived by combining physiological cascading and
cellular processes. The model goes beyond compartmental models in
that we have included both the nonlinear absorption of IDL and LDL
by cells, and a feedback mechanism by which the cell regulates the
number of its LDL receptors based on the concentration of
intracellular cholesterol IC.

This low-dimensional, nonlinear model shows bistability and
hysteresis between a low and a high cholesterol state. Our
bifurcation analysis of the key control parameters ($u_V$ and
$d_{IC}$) can be related to physiologically meaningful processes:
diet and statin medication in the case of $u_V$; reverse
cholesterol transport through HDL in the case of $d_{IC}$. We have
checked that the observed behaviour is relatively robust with
respect to all other parameters. An important outcome of our
sensitivity analysis is that the most robust feature in the low
cholesterol state is the concentration of intracellular
cholesterol $[IC]$, while the plasma concentrations $[IDL]$ and
$[LDL]$ can vary widely. This indicates that plasma cholesterol is
not a tightly controlled variable in the system, implying that
very different values of plasma $[LDL]$ can be compatible with
viable cellular function. It is conceivable, however, that the
high levels of $[LDL]$, innocuous at the cellular level, could
lead to the initiation of disease-linked processes at the
physiological level.

We have also obtained estimates for the characteristic time scales
governing the dynamics of the model in the low and high
cholesterol states. These time scales account for the dynamical
responses of the system to forced input oscillations with very
different periods. The model shows dynamical behaviour which could
be used to diagnose the state of the system if repeated
measurements of plasma and/or cellular cholesterol with the
correct timing and periodicity are performed, as indicated by
Figs.~\ref{graph6}~and~\ref{100days}. If measurements are too
sparse, the onset of permanent high cholesterol levels might go
undetected, with the possibility of a difficult return to lower
levels due to hysteretic effects. Measurements taken periodically
on the order of hours or months could provide complementary clues
to the state of the system. If we assume that periodic inputs to
the system are important, measuring the relative variability of
LDL and cellular cholesterol (possibly through skin tests) could
lead to markers for the detection of the intrinsic regime of the
system.

Our dynamical model is directly derived from the underlying
physiological and cellular processes. However, some of our
approximations are crude and several directions for improvement
could be pursued. The level of the plasma cholesterol on the high
cholesterol branch ($[VLDL]_{eq}+[IDL]_{eq}+[LDL]_{eq}$) is
unphysiologically high in the current model. This suggests the
presence of other physiological mechanisms of elimination that
have not been included in the model. Another extension would
include time delays~\citep{LDEIV98} to account for mass transfer
(e.g., the delayed intake of IDL and LDL via LR-mediated
endocytosis) and control mechanisms (e.g., the delayed genetic
regulatory loop in the synthesis of LR). Furthermore, the model
could be extended to include explicitly the regulatory and
synthetic pathways involving the intracellular LDL receptors. It
would also be important to distinguish between hepatic and
non-hepatic cells, as hepatic cells play a more central role in
lipoprotein metabolism. In particular, the rate of lipoprotein
uptake in liver cells is different; increased uptake of LDL in
liver cells lowers the production rate of VLDL (i.e., $u_V$)
through an additional regulating feedback; and statins reduce the
inhibiting effect of $[IC]$ on the synthesis of LR in liver cells
only.

Especially important would be to extend our description of reverse
cholesterol transport to include $[HDL]$ explicitly as a variable.
Indeed, $20\%$ of the total plasma cholesterol in healthy
individuals ($40 \,mg/dl$) is carried by HDL. Our preliminary work
in this direction indicates that the number of nonnegative
equilibria remains unchanged in the enlarged system that includes
$[HDL]$, and the qualitative features of the system should hold.
Because the relationship between plasma $[LDL]$ and $[HDL]$ is of
major importance in clinical trials, this would be a crucial
element to enhance the applicability of the model.

Further modelling of lipoprotein metabolism could bring light to
several open questions. To name but a few, it is still unknown why
the plasma cholesterol concentrations in adult humans are so
unnecessarily high~\citep{JM77}; what the exact effects of statins
in the metabolic system are~\citep{MHD01}; and which of the
processes could be best targeted to reduce plasma LDL
concentrations.



\appendix{\bf{Appendices}}
\section{Derivation of the feedback term}\label{appendix1}

\begin{figure}[hb]
\vspace*{-2cm}
\centering
\includegraphics[width=1.2\textwidth]{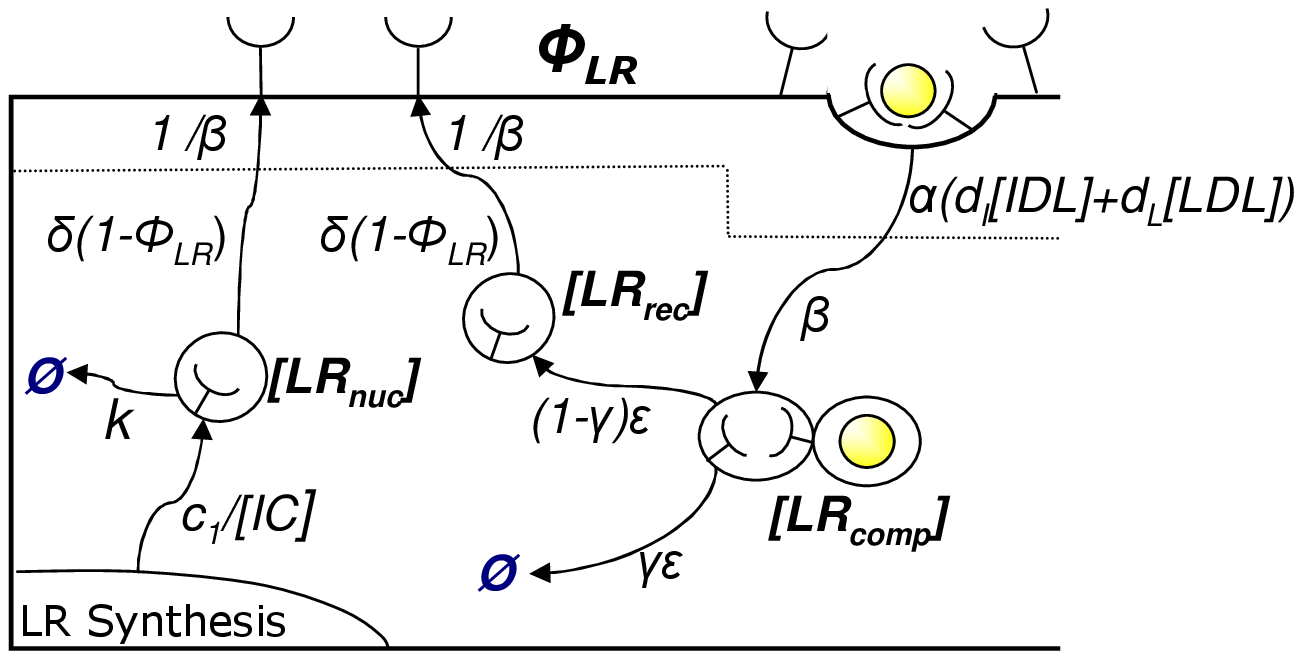}
\vspace*{-6.5in} \caption[\textbf{Schematic description of the
feedback term.}]{\textbf{Schematic description of the feedback
term.} {\small An expanded view of the portion of
Figure~\ref{fig7} that is related to the processes involved in the
synthesis and recycling of LDL receptors inside the cell. If we
assume steady state for the intracellular LR compartments, and
that the turnover of nucleus-synthesised LR ($k$) is large
compared to the probability of its insertion into the membrane
($\delta$), these processes lead to Equation~(\ref{full4}). The
factor $\beta$ performs the appropriate dimensional conversions to
account for the fact that $\phi_{LR}$} is a non-dimensional
surface occupancy fraction. Note also that $\gamma$ is the
fraction of non-recycled LDL receptors. All the other constants
are rate constants with the appropriate
units.}\label{fig_appendix}
\end{figure}

The feedback term in Eq.~(\ref{full4}) follows from a more
detailed (but still simplified) derivation that includes the three
compartments of intra-cellular LDL receptors (complexed, recycled,
nucleus-synthesised) as shown in Figure~\ref{fig_appendix}. The
kinetic rate equations are:
\begin{align}
\frac{d[LR_{comp}]}{dt}&=\beta \alpha \left(d_I[IDL]+d_L[LDL] \right )\phi_{LR} - \epsilon  [LR_{comp}]\label{app1}\\
\frac{d[LR_{rec}]}{dt}&=\epsilon (1- \gamma) [LR_{comp}] - \delta [LR_{rec}] \left(1-\phi_{LR}\right)\label{app2}\\
\frac{d[LR_{nuc}]}{dt}&= \frac{c_1}{[IC]} - k [LR_{nuc}] - \delta [LR_{nuc}] \left(1-\phi_{LR}\right)\label{app3}\\
\frac{d\phi_{LR}}{dt}&=-\alpha \left(d_I[IDL]+d_L[LDL]
\right)\phi_{LR}+ \frac{\delta}{\beta} ([LR_{rec}] + [LR_{nuc}])
\left(1-\phi_{LR}\right). \label{app4}
\end{align}
If we assume steady-state conditions: $\frac{d[LR_{comp}]}{dt} =
\frac{d[LR_{rec}]}{dt} = \frac{d[LR_{nuc}]}{dt} =0$, and that $k
\gg \delta$, then Eq.~(\ref{full4}) follows with $b=\alpha \gamma$
and $c=\delta c_1/ \beta k$.

\end{document}